\documentclass[8.5pt,twoside,twocolumn]{article}
\usepackage[super,sort&compress,comma]{natbib} 
\usepackage{mhchem}
\usepackage{times,mathptm}
\usepackage{sectsty}
\usepackage{balance} 

\usepackage{graphicx} 
\usepackage{lastpage}
\usepackage[format=plain,justification=justified,singlelinecheck=false,font=small,labelfont=bf,labelsep=space]{caption} 
\usepackage{fancyhdr}
\usepackage{amssymb,amsmath}
\usepackage[usenames,dvipsnames]{color}
\usepackage{balance}

\begin{document}

\thispagestyle{plain}
\fancypagestyle{plain}{
\fancyhead[L]{\textsf{\textcolor{Gray}{PAPER}}}
\fancyhead[R]{\textsf{\textcolor{Gray}{www.rsc.org/softmatter}~~~\textbar~~~Soft Matter}}
\renewcommand{\headrulewidth}{1pt}}
\renewcommand{\thefootnote}{\fnsymbol{footnote}}
\renewcommand\footnoterule{\vspace*{1pt}%
\hrule width 3.4in height 0.4pt \vspace*{5pt}} 
\setcounter{secnumdepth}{5}

\makeatletter 
\def\subsubsection{\@startsection{subsubsection}{3}{10pt}{-1.25ex plus -1ex minus -.1ex}{0ex plus 0ex}{\normalsize\bf}} 
\def\paragraph{\@startsection{paragraph}{4}{10pt}{-1.25ex plus -1ex minus -.1ex}{0ex plus 0ex}{\normalsize\textit}} 
\renewcommand\@biblabel[1]{#1}            
\renewcommand\@makefntext[1]%
{\noindent\makebox[0pt][r]{\@thefnmark\,}#1}
\makeatother 
\renewcommand{\figurename}{\small{Fig.}~}
\sectionfont{\large}
\subsectionfont{\normalsize} 

\fancyfoot{}
\fancyfoot[LO,RE]{\footnotesize \textsf{\textcolor{Gray}{This journal is \copyright The Royal Society of Chemistry 2010}} }
\fancyfoot[RO]{\footnotesize{\sffamily{\textit{Soft Matter}, 2010, \textbf{X},
      1--\pageref{LastPage} ~\textbar \hspace{2pt}\thepage}}}
\fancyfoot[LE]{\footnotesize{\sffamily{\thepage~\textbar \textit{Soft Matter}, 2010, \textbf{X}, 1--\pageref{LastPage}}}}
\fancyhead{}
\renewcommand{\headrulewidth}{1pt} 
\renewcommand{\footrulewidth}{1pt}
\setlength{\arrayrulewidth}{1pt}
\setlength{\columnsep}{6.5mm}
\setlength\bibsep{1pt}

\twocolumn[
  \begin{@twocolumnfalse}
\noindent\Large{\textbf{Universal two-step crystallization of DNA-functionalized nanoparticles}}
\vspace{0.6cm}

\noindent\large{\textbf{Wei Dai,\textit{$^{a}$} Sanat K. Kumar,\textit{$^{b}$} and
Francis W. Starr$^{\ast}$\textit{$^{a}$}}}\vspace{0.5cm}

\noindent\textit{\small{\textbf{Received 7th June 2010, Accepted 12th
      August 2010
}}}

\noindent \textbf{\small{DOI: 10.1039/b000000x}}
\vspace{0.6cm}

\noindent \normalsize{ We examine the crystallization dynamics of nanoparticles reversibly
  tethered by DNA hybridization. We show that the crystallization
  happens readily only in a narrow temperature ``slot,'' and always
  proceeds via a two-step process, mediated by a highly-connected
  amorphous intermediate.  For lower temperature quenches, the dynamics
  of unzipping strands in the amorphous state is sufficiently slow that
  crystallization is kinetically hindered.  This accounts for the
  well-documented difficulty of forming crystals in these systems.  The
  strong parallel to the crystallization behavior of proteins and
  colloids suggests that these disparate systems crystallize in an
  apparently universal manner. }  \vspace{0.5cm}
\end{@twocolumnfalse}
  ]

\footnotetext{\textit{$^{a}$~Department of Physics, Wesleyan
  University, Middletown, Connecticut 06459, USA Fax: +1 860 685 2031;
  Tel: +1 860 685 2044; E-mail: fstarr@wesleyan.edu}}
\footnotetext{\textit{$^{b}$~Department of Chemical Engineering,
  Columbia University, New York, New York 10027, USA }}

\section{Introduction}

The use of DNA as a programmable linking agent is a practical,
``bottom-up'' approach to materials
design~\cite{niemeyer,seeman2,condon06,ge10}.  One starts from
``molecules'', consisting of nanoparticles (NP) functionalized by
multiple single stands of DNA (ssDNA). When the DNA on adjacent
particles hybridize to form double-stranded DNA (dsDNA) the particles
are physically linked, potentially leading to the formation of complex
structures~\cite{mirkin,alivisatos96}.  If one can specify the number
and orientation of the ssDNA attached to the NP, it is possible to
control the local geometry of the networks, which in turn may control
the geometry of higher-order structures~\cite{glotzer07}. This bottom-up
approach can result in a precision hard to achieve by molecular
nano-fabrication, with promising future applications in optical and
electrical materials~\cite{crocker08, seeman2}. While there have been
some recent successes creating crystalline ordered arrays of DNA-linked
NP~\cite{crocker09, mirkin08, gang08,gang09, kim06, macfarlane09}, the
formation of regularly ordered structures has proved challenging. More
frequently, the NP assemble into disordered
aggregates~\cite{park-mirkin04, crocker05, nykypanchuk-gang07, maye06,
  maye07, park01, eiser08}.  Therefore, we aim to better understand the
dynamical pathways the system must follow in order to crystallize, as
well as the mechanisms that hinder ordering, with the ultimate goal of
avoiding kinetic bottlenecks.

To put these difficulties in the context of more traditional materials,
experiments on colloidal particles, which isotropically interact with
each other, are well-known to exhibit a two-step crystallization
mechanism. In these cases, the gas-liquid coexistence curve is
metastable relative to the gas-solid coexistence curve. Quenching these
colloidal system inside the gas-liquid coexistence region results in (i)
a phase separation into a high-density liquid, followed by (ii) crystal
nucleation within this high-density droplet. Quenching the system to
very low temperature causes it to simply form disordered gels,
kinetically hindering crystal formation. There is considerable
theoretical and simulation evidence for this scenario~\cite{hobbie98,
  lkd09, filobelo05, vekilov2,galkin00b,galkin00a,vekilov1, garetz02,
  bonnett03,qian04, lu08}.  Similar ideas have also been proposed for
proteins, and it is now believed that there is a crystallization
temperature ``slot'' outside which crystallization does not occur
\cite{vekilov2,galkin00b,galkin00a,vekilov1}. An alternate mechanism for
clustering preceding crystallization has also been proposed for the case
of proteins interacting through strong, but patchy interactions.  It has
been argued that self-assembly, driven by highly specified local
geometry imposed by the bonding sites, can create a locally high-density
region which enhances the formation of nuclei, even in the absence of
the thermodynamic drive for phase separation. A two-step mechanism,
mediated by this self-assembled amorphous state, is thus another pathway
by which proteins can crystallize~\cite{lkd09}. While the factors
controlling the crystallization of colloids and proteins (or patchy
colloids) may be system specific, we stress that crystallization in
these soft matter systems always seems to follow a two-step kinetic
scheme, with an amorphous, highly-connected phase serving as a kinetic
intermediate.

Motivated by this apparent universality in these systems, we examine the
crystallization of DNA-linked NP, where the formation of clusters is
controlled by DNA hybridization. We show that, in spite of the
significant differences in the physical connectivity between this case
and previously examined situations ({\it i.e.}, colloids and proteins),
the crystallization of NPs linked by DNA tethers also follows a two-step
process: initially there is a cluster of linked particles without any
crystalline order. This process is facilitated either by self-assembly
of the nanoparticles, or by phase separation, depending on the region of
parameter space explored. We show that the persistence of this amorphous
state grows very rapidly on cooling, resulting in a very narrow
crystallization temperature slot.  We thus argue that the
crystallization of such ssDNA tethered NPs follow the same universal
behavior as found for colloids, proteins and other patchy colloids.

\section{Modeling}
In our simulations, each NP is grafted with 6 chains in an octahedral
symmetry. This orientation will naturally lead these NP to crystallize
into a simple cubic structure. We use an effective potential model
developed to capture the base-pair selectivity between two ssDNA based
on the nucleotides' identity (A, T, C, or G), and the bonding
specificity which allows only one bond to each
base~\cite{ss06,lts07}. The effective potential between two ssDNA
depends only on the intermolecular separation and their relative angular
orientation. The parameters of the effective potential are obtained by a
systematic coarse-graining of a more detailed model for the DNA
interactions, and it has been verified that the coarse-grained model
quantitatively reproduces the behavior of the more explicit
model~\cite{hlss08, lts07}.  We study this model via Monte Carlo (MC)
simulations in the canonical ensemble (fixed number of particles $N$,
volume $V$, temperature $T$). For a randomly chosen NP, we perform three
sequential attempts -- (i) NP translation without rotation, (ii) NP
rotation without translation, and (iii) a combined NP translation and
rotation. We define one Monte Carlo step (MCS) as $N$ such attempts,
where $N$ is the total number of NP.  Note that the dominant
interactions in this system are between the bonding arms; the
short-ranged NP core repulsion plays relatively little role.  There are
no explicit solvent interactions, so we cannot directly capture effects
such as salt concentration.  We used reduced units, as defined in
ref.~\cite{ss06}. To improve the statistics of our results, we average
over 5-15 independent runs for each system.

The formation of a crystal or liquid phases in this system is driven by
the DNA base pairing alone, resulting in very low density phases. The
addition of significant attractions between NP cores -- which might be
expected in some experimental systems -- can result in much higher
density phases, where the density is determined by the packing of the NP
cores.  Such phases are analogous to those found for isotropically
interacting colloids.  We do not consider this more complicated
situation in this paper, and defer it to future research.

Previously, ref.~\cite{dhss09} examined the phase diagram for this
model, and showed that it exhibits polymorphous phase behavior with at
least six distinct crystal phases. The lowest density crystal, crystal
I, consists of a simple cubic (SC) lattice that reflects the octahedral
symmetry of the functionalized NP. The length of the DNA connections
between NP leaves ample space, allowing the cubic order to repeat itself
as a hierarchy of interpenetrating cubic lattices. Accordingly, crystal
II consists of two interpenetrating SC lattices, crystal III has three
interpenetrating SC lattices, {\it etc}.  Experimentally, achieving this
interpenetration may be challenging, due to repulsions from the charged
DNA backbone.  However, with a proper choice of solution, electrostatic
interactions beyond a few nm between DNA can be effectively
screened~\cite{rpbz06}. Accordingly, functionalizing strands should be
long enough to open the network structure and thereby minimize
electrostatic repulsions.

\begin{figure}
\begin{center}
\includegraphics[width=8.8cm]{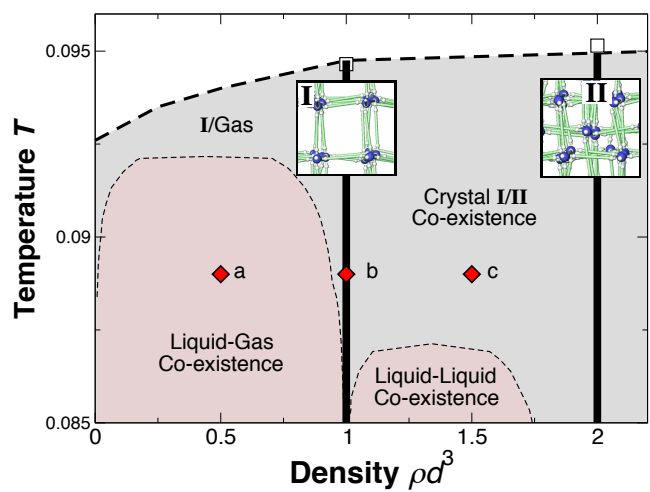}
\end{center}
\caption{ The phase diagram of the octahedrally functionalized
  NP~\cite{dhss09}. Solid vertical lines are the densities of crystal I
  and II, and the images show simulation snapshots of the local order of
  each crystal; in these snapshots, the blue spheres represent the core
  NP units, and DNA bonds between NP are represented by the green
  connections.  The grey regions of the phase diagram indicate
  crystal-gas or crystal-crystal coexistence.  The crystal melts on
  heating to at the bold, dashed line.  The faded-red regions are the
  metastable amorphous phase separation regions.  The estimated phase
  boundaries for the gas-liquid ($0<\rho d^3 <1$) and liquid-liquid
  ($1<\rho d^3 <2$) phase transitions are indicated by the light dashed
  lines.  The red diamonds indicate the state points where we quench to
  study the crystallization dynamics -- one point in the phase amorphous
  separation region (a), and two in assembly dominated regimes (b) and
  (c).  These letters correspond to those used in
  Fig.~\ref{fig:time_evolve}. }
\label{fig:phase_boundary}
\end{figure}

Fig.~\ref{fig:phase_boundary} shows the melting temperature for crystal
I and II and the metastable amorphous phase boundaries; the phase
boundary of the metastable amorphous phases is estimated by
extrapolating the observed phase boundaries of NP functionalized by 3,
4, or 5 DNA strands in ref.~\cite{dhss09}.  Parallel to the crystal
polymorphism, there is liquid state polyamorphism -- that is, this
system exhibits the unusual feature of multiple thermodynamically
distinct liquid states in a pure system~\cite{pses92,pgam97,ms98review}.

We focus primarily on quenches to $T=0.089$ where we empirically find
that crystallization proceeds most readily. We also consider other $T$
to test how the rate of crystallization varies.  We investigate systems
at $\rho d^3=0.5$, $\rho d^3=1$, and $\rho d^3=1.5$, where $\rho d^3$ is
the scaled density so that a single SC lattice has density $\rho
d^3=1$~\cite{dhss09}. We keep $V$ fixed, so these three densities
correspond to $N=500,\ 1000,\ 1500$, respectively. The system at $\rho
d^3=0.5$ is located in the amorphous phase separation region, whereas
other two systems at $\rho d^3=1$, and $\rho d^3=1.5$ are outside of any
amorphous-amorphous phase transition. These three systems allow us to
separately address the role of single versus double-interpenetrating
crystals, and spinodal-assisted versus assembly-driven crystallization
processes.

\section{Results}

Since we wish to track the dynamics of crystallization, we need to
separately identify the formation of clusters and the degree of
crystallinity of those clusters.  First, we evaluate the evolution of
the cluster size directly from the number of bonded NP and calculate the
weighted mean cluster size
\begin{equation} s(t) = \frac{ \sum_{\rm clusters} n^2
  P(n)} {\sum_{\rm clusters} {n P(n)}} 
\end{equation}
where $n$ is the size of a cluster at some time $t$, and $P(n)$ is the
probability of finding a cluster of size $n$.  Here, a bond is
determined by the linkage of the DNA strands connecting NP.

\begin{figure}[h!t]
\begin{center}
\includegraphics[width=8.8cm]{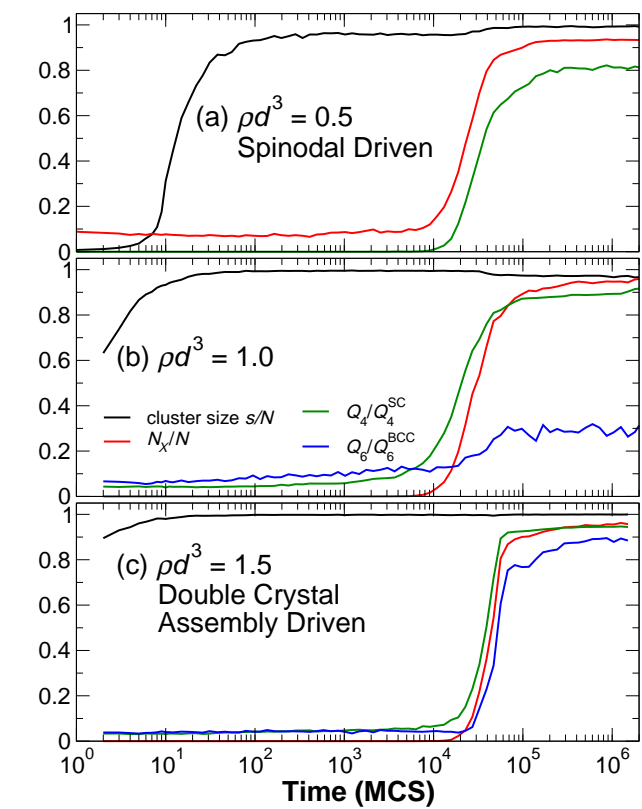}
\end{center}
\caption{ The time evolution of the clustering and crystallization
  process at (a) $\rho d^3=0.5$, where spinodal decomposition promotes
  clustering; (b) $\rho d^3=1$, the density of the single SC crystal,
  where DNA hybridization drives clustering; (c) $\rho d^3=1.5$, where there
  is a phase separation of the single and double interpenetrating
  crystals.  For each density, we show the fractional cluster size
  $s/N$, the fraction of crystal-like particles $N_X/N$, and the
  normalized average orientation $Q_4/Q_4^{\rm SC}$. For $\rho d^3=1$
  and $1.5$, we also show $Q_6/Q_6^{\rm BCC}$ to determine whether the
  formation of two interpenetrating crystals are simultaneous or occur
  step-wise. }
\label{fig:time_evolve}
\end{figure}

Following the approach in ref.~\cite{wrf96}, we can identify
crystal-like regions using an orientational order parameter $Q_l$,
defined by a sum over spherical harmonic functions $Y_{l,m} (\hat{\bf
  r})$ using the unit vectors $\hat{\bf r}$ defined by the bonded
neighbors. Among the various choices for the degree $l$ of $Y_{l,m}
(\hat{\bf r})$, $l=4$ provides the strongest signal for the expected
cubic symmetry; specifically, $Q_4 = 0.764$ for an ideal SC lattice. To
capture the locally SC structure in case of two separate lattices, we
consider only bonded neighbors when calculating $Q_4$, excluding
unbonded neighbors, which might have small separation but belong to a
distinctly different bonded network.  Additionally, we compute $Q_6$ for
nearest neighbors to identify possible body-centered-cubic (BCC)-like
structures of non-bonded units that might arise due to interpenetration
at $\rho d^3=1, 1.5$. For amorphous systems, $Q_l = 0$ in the
thermodynamic limit, so that one can immediately distinguish crystal
from amorphous systems.

While $Q_4$ and $Q_6$ are useful to identify global crystallinity in a
given configuration, they are less helpful in identifying the presence
of local crystal regions embedded in a larger amorphous cluster.
Identification of these locally crystalline regions is necessary to
track if crystals can assemble directly, or if crystals only form
following a connected amorphous intermediate state.  Following
ref.~\cite{wrf96}, we identify crystal-like particles using a local
invariant $q_4(i)$ for each individual particle, and the corresponding
complex vector ${\bf q_4(i)}$.
A particle is said to be crystal-like if it has a minimum number of
neighbors with crystal-like connections.  Bonded particles $i$ and $j$
are said to have a crystal-like connection if the vector dot product
${\bf q_4(i)} \cdot {\bf q_4(j)}$ exceeds a threshold value.  By
comparing the distribution of dot product values of amorphous systems
with well-crystallized systems, we find that a threshold value ${\bf
  q_4(i)} \cdot {\bf q_4(j)} \geq 0.95$ for the dot product identifies
more than $90\%$ of bonds in crystal state, and only misidentifies less
than $2\%$ of the bonds of the amorphous system as crystal-like at $\rho
d^3=0.5$. Similar precision is found at $\rho d^3=1$ and $\rho
d^3=1.5$. Finally, analysis of these systems shows that we reliably
define a crystal-particle if it has at least three crystal-like bonded
neighbors.  We use this as our criterion to label an NP as part of a
crystal.

We examine the evolution of the crystallizing systems by evaluating the
fractional cluster size $s/N$, the fraction of crystal-like
nanoparticles $N_X/N$, and the normalized average orientation
$Q_4/Q_4^{\rm SC}$, so that all quantities vary over the range $[0,1]$
(Fig.~\ref{fig:time_evolve}). The density $\rho d^3=0.5$ allows us to
examine crystallization to a single network in the presence of an
amorphous phase separation, similar to the case of colloidal systems.
For comparison, density $\rho d^3 = 1.0$ follows crystal formation
driven only by the assembly of DNA links.  Finally, for $\rho d^3=1.5$,
self-assembly (without phase separation) drives the formation of a
higher density state so that the system crystallizes into a combination
of a single cubic lattice and a second higher density interpenetrating
lattice. Accordingly, we can determine if the pathway for
crystallization for interpenetrating networks differs from that for a
single network, and compare spinodal driven versus self-assembly driven
clustering.

Figure 2 shows that, after quenching from high $T$ to $T=0.089$ (just
below the hybridization temperature for the DNA), we find that all
systems rapidly undergo a condensation from an unbonded state to a
large, bonded amorphous cluster. Both $Q_4$ and $N_X$ remain small for
$t \lesssim 10^4$, demonstrating the amorphous nature of the
cluster. The ordering process only occurs much later, at $t \approx
10^4$ when $Q_4$ and $N_X$ rise sharply within a narrow window of
time. Since the ordering happens over a relatively narrow time window,
the global $Q_4$ is itself an indication of crystal formation. Note that
the fraction $N_X/N$ never reaches one, since there are always surface
particles of the crystal that will not be identified as crystal-like.
These results establish that, for each state point considered, the
system first forms a highly connected amorphous phase, from which a
crystal nucleates and grows.  We refer to this as the two-step process
of crystallization.

Having established the two-step nature of crystallization, we next wish
to determine if the intermediate amorphous phase can be considered a
metastable equilibrium, and how this impacts the difficulty of the
eventual crystallization.  To do so, we examine the crystallization
dynamics as a function of the quench depth by evaluating the time
$\tau_X$ needed for the crystal to nucleate and the internal relaxation
time $\tau_\alpha$ of the amorphous intermediate.  We define $\tau_X$ by
the time when 10\% of the particles are designated as crystal, since the
crystallization process appears irreversible at this fraction; an
alternate criterion will change the value of $\tau_X$, but not the $T$
dependence.  We define $\tau_\alpha$ for the amorphous phase from the
relaxation time of the coherent intermediate scattering function
$F(q,t)$, evaluated at the wave vector $q$ corresponding to the bonding
distance between NP; we choose this wave-vector since it captures the
slowest relaxation (apart from $q\rightarrow 0$).  The determination of
$\tau_\alpha$ is complicated due the the aging of the amorphous state
following the temperature quench.  To limit aging effects, we wait for
the largest possible time to begin calculating $F(q,t)$ that will still
allow $F(q,t)$ to decay to zero prior to crystallization.

\begin{figure}
\begin{center}
\includegraphics[width=8.8cm]{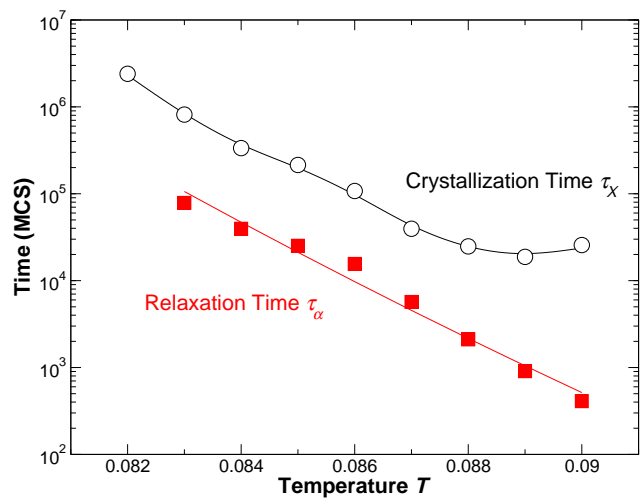}
\end{center}
\caption{ Time-temperature-transformation diagram showing the
  temperature dependence of the crystallization time $\tau_X$ and
  amorphous intermediate relaxation time $\tau_\alpha$.  The minimum of
  $\tau_X$ shows there is only a narrow slot where crystallization
  readily proceeds.}
\label{fig:crystal-time}
\end{figure}

Figure~\ref{fig:crystal-time} shows the time crystallization time
$\tau_X$ and the amorphous relaxation time $\tau_\alpha$ for density
$\rho d^3 = 0.5$, which have been averaged over 10 independent
trajectories.  Such a plot is commonly referred to as a
``time-temperature-transformation''
diagram~\cite{debenedetti-book,mka83}.  We find that $\tau_\alpha$ is
significantly smaller than $\tau_X$ so that the clustered state can
reach a metastable equilibrium prior to crystallization. Hence the first
step toward crystallization also includes the equilibration of the
metastable state.  On cooling, $\tau_\alpha$ increases rapidly, since
the lifetime of dsDNA pairs grows quickly, thereby hindering relaxation.
In contrast, $\tau_X$ initially decreases on cooling, expected since the
predicted barrier to crystallization decreases on cooling from classical
nucleation theory~\cite{debenedetti-book}.  However, on further cooling,
$\tau_X$ rapidly increases, as it become dominated by the slow
relaxation of $\tau_\alpha$.  From a physical perspective, the
persistence of DNA base-pair bonds prevents unzipping on a reasonable
time scale, and thus the amorphous cluster cannot reorganize to ``find''
the crystal state, resulting in a kinetically dominated process.
Accordingly, there is a very narrow slot that must be found for
successfully nucleating the crystal state.  The minimum, or ``nose'', in
the crystallization time is ubiquitous in supercooled
liquids~\cite{debenedetti-book,mka83,kbjrr96,ymss02}.  If the system is
cooled below $T_{\rm nose}$ in a time less than $\tau_{\rm nose}$, the
system will not have adequate time to crystallize. Accordingly, the nose
defines a critical cooling rate $R = (T_M- T_{\rm nose})/\tau_{\rm
  nose}$, where $T_M$ is the melting
temperature~\cite{debenedetti-book}; cooling faster than this rate will
prevent crystallization in all cases, and should therefore be avoided.

For the highest density $\rho d^3=1.5$, we wish to further determine if
the formation of a double network occurs simultaneously with, or after
the formation of a single SC network.  To test this, we check for the
presence of BCC order in the lattice using $Q_6$ (since it is more
sensitive to BCC order) and compare its evolution to $Q_4$.  For the
calculation of $Q_6$, we use spatial separation, rather than bonds, to
determine neighbors, since the units comprising the BCC structure are
actually unbonded neighbors in a separate cubic lattice.  These unbonded
neighbors are separated by a distance {\it less} than the bonding
distance.  Choosing a cutoff separation of $\sqrt{3}/2\, d \approx 0.866
\, d$ (ratio of BCC to FCC lattice spacing) effectively excludes cubic
bonded neighbors, and includes most neighbors that should have BCC
order.  We actually use a slightly less restrictive definition, and
include neighbors up to a distance $0.92 \, d$, since the positions of
interpenetrating particles are not rigidly fixed; this cutoff is still
small enough that it excludes the vast majority of bonded neighbors.

For reference, we first examine density $\rho d^3=1.0$ where there
should be no interpenetration, so that we know to what degree $Q_6$
might give a false signal of interpenetration.
Fig~\ref{fig:time_evolve}(b) shows that $Q_6/Q_6^{\rm BCC} \approx 0.3$
for the final SC crystal, so that we have little false signal of
interpenetration.  The small value of $Q_6/Q_6^{\rm BCC}$ can be largely
attributed to our choice of the cutoff distance used to determine
neighbors, since if we included the bonded neighbors of the SC lattice
we would expect $Q_6/Q_6^{\rm BCC} = 0.693$ (for a perfect SC lattice).
Applying this metric for the case with interpenetration ($\rho
d^3=1.5$), Fig.~\ref{fig:time_evolve}(c) shows that $Q_6/Q_6^{\rm BCC}$
captures the eventual interpenetration.  However, the growth of $Q_6$
slightly lags behind that of of $Q_4$. Thus, the formation of a single
cubic lattice appears to slightly precede the interpenetrating
structure.  Note that the asymptotic value of $N_X/N$ for $\rho d^3=1.5$
is slightly larger than for the lower densities.  This is because we use
a larger $N$, and thus the surface-to-volume ratio is smaller, and so
the surface effect on $N_X/N$ is smaller.

For both densities, note that $s$ slightly increases upon
crystallization.  Presumably, the long ranged order allows for the
formation of some additional bonds that were dangling in the clustered,
but disordered state.  For $\rho d^3=1.5$, $s$ has a very small but
noticeable {\it decrease} immediately prior to ordering. This can be
understood by the fact that the formation of two interpenetrating SC
lattices requires a separation of two lattices; therefore, there must be
a temporary breaking of amorphous bonds between locally cubic lattices
before ordering, which temporarily decreases $s$.

\begin{figure}
\begin{center}
\includegraphics[width=8.8cm]{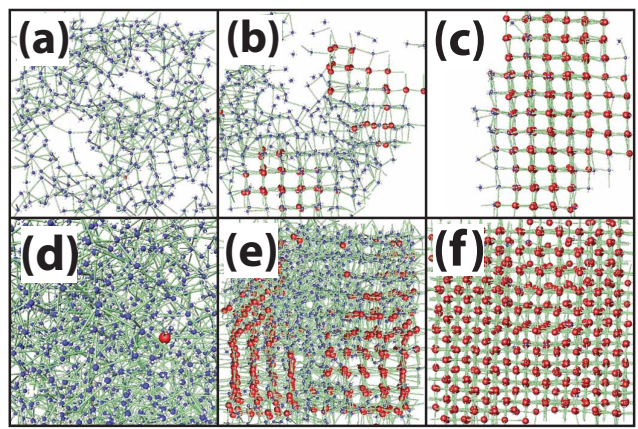}
\end{center}
\caption{Visualization of the system at three points during
  crystallization. (a)--(c) are at density $\rho d^3=0.5$. Specifically,
  (a) is system after clustering but prior to ordering; (b) is during
  the ordering; (c) is after ordering. (d)--(f) represent the same time
  progression but at density $\rho d^3=1.5$. The crystal-like particles
  are colored in red with a slightly larger size.}
\label{fig:visualization}
\end{figure}

We directly visualize the crystallization of two systems ($\rho d^3=0.5$
and $\rho d^3=1.5$) at three important points: (i) clustering prior to
ordering; (ii) the ordering process, and (iii) after ordering
(Fig.~\ref{fig:visualization}). The crystal-like NP are colored in red
with a slightly larger radius. From Fig.~\ref{fig:visualization} (b) and
(e) we can clearly see the interface between crystal and amorphous
phases.  Fig.~\ref{fig:visualization} (b) has two separate SC lattices
and evidence of multiple nucleation cores.

\section{Discussion and Conclusions}

The observed sequence of clustering via DNA links followed by ordering
for {\it all} densities confirms that the crystallization dynamics for
the DNA-linked NP follows a conventional ``two-step'' pathway of
crystallization -- even in the case where a double-interpenetrating
network must form.  Depending on density, the amorphous intermediate is
driven either through metastable phase separation or DNA driven
assembly. Additionally, the intermediate amorphous state has a rapidly
growing lifetime on cooling.  Thus, the same framework used to
understand protein and colloid crystallization kinetics also applies to
this more unusual material.

We compare our results with the recent experimental studies of Mirkin
and co-workers~\cite{macfarlane09}, where they report a 3-stage
crystallization process for uniformly DNA-coated NP.  In that study, the
NP initially form small amorphous aggregates, and these clusters
separately evolve crystallinity at stage two. The ordered but dispersed
clusters eventually coalesce into a large final crystal lattice. As
noted by these authors, this last stage of crystallization is probably
driven by Ostwald ripening, a mechanism that is relatively well
explored.  Accordingly, these experiments also fit within the general
framework of the two-step process, since the crystallinity of small
clusters evolves from the amorphous aggregates.  To further compare to
these experiments, we also checked if very slow ripening might occur in
our simulations under different thermodynamic conditions.  Indeed,
simulations at lower density ($\rho d^3=0.3$) evolve crystallinity in
small region within the amorphous cluster.  The subsequent growth of
this crystal is so slow, that we are not able to complete the
crystallization in the computational time frame of the simulation.  This
slower scenario for crystal growth is likely the same as that observed
in ref.~\cite{macfarlane09}.  Accordingly, the rate of the ``second''
step in the process (crystal growth) can vary significantly depending on
the state point chosen.  We should note a potentially important
difference between our simulations and that of the experiments of
ref.~\cite{macfarlane09}.  For the uniformly coated NP used
experimentally, the local NP binding is not predisposed to reflect the
order of the eventual crystal state.  In contrast, our 6-armed units
bind in such a way to readily reflect SC symmetry.  We expect that the
coarsening to the eventual crystal state should occur more readily in
the case where the crystal order is reflected in the symmetry of local
bonding.  Nonetheless, since the experimental system and the simulated
model behave similarly, we conclude that the two-step crystallization is
the apparently universal pathway followed by these systems.

Experiments using nanoparticles functionalized with many strands of DNA
(as opposed to a small number of strands) present an additional
potential barrier to crystallization.  As the strand density increases,
it has been experimentally found that the hybridization transition
becomes increasingly sharp~\cite{jwlms03}.  This behavior has also been
argued for theoretically~\cite{ge10}.  Narrowing the hybridization
window will also result in narrowing the crystallization temperature
slot -- making the formation of ordered structures even more
challenging.  Accordingly, it may be valuable to work with NP
functionalized by a small number of strands.  These limited
functionality systems also have the advantage that they lend themselves
more readily to a theoretical description~\cite{hss10}.

Since there are no NP attractions in our model (other than indirectly
via DNA linking), the gas-liquid phase separation is driven by the
self-assembling DNA-hybridization. If, in addition to the hybridization,
there were isotropic NP attractions, the system might be able to form a
much higher density droplet controlled by the packing of the NP cores.
This could result in a crystallization by hybridization that would
actually dramatically {\it decrease} the density relative to the
amorphous cluster, since DNA links will serve to open the structure,
qualitative similar to the crystallization of ice from water.  Such an
intermediate not created by DNA base pairing might offer different
pathways to creating crystals, without the kinetic traps that are
encountered experimentally.  This is another possible avenue to pursue
to experimentally facilitate the crystallization process.

The highly specified DNA orientation in our model clearly helps to build
in self-assembly of the higher-order crystal structure.  However, this
is not a sufficient condition to guarantee the ready formation of
crystals.  If that were the case, a previous study of NP decorated with
ssDNA in a tetrahedral orientation~\cite{hlss08} should have discovered
the spontaneous formation of diamond lattices, but instead found only
amorphous gels.  Hence, significant work remains to understand how to
best design a desired higher order structure from relatively simple
building blocks.
\bigskip

\section{Acknowledgments}
  We thank C.W.~Hsu, F.~Sciortino, and F.~Vargas for helpful
  discussions. We acknowledge the NSF for support under grant number
  DMR-0427239, and we thank Wesleyan University for computer time
  supported by the NSF under grant number CNS-0959856.

\footnotesize{

\providecommand*{\mcitethebibliography}{\thebibliography}
\csname @ifundefined\endcsname{endmcitethebibliography}
{\let\endmcitethebibliography\endthebibliography}{}

}
\balance

\end{document}